\documentclass[twocolumn,showpacs,preprintnumbers,amsmath,amssymb]{revtex4}
\usepackage{graphicx}
\usepackage{dcolumn}
\usepackage{bm}
\newif\ifappendixon
\begin{document}

\preprint{APS/123-QED}

\title{Manipulation of optical solitons in Bose-Einstein condensates}
\author{A.V. Rybin}

\affiliation{Department of Physics, University of Jyv\"askyl\"a PO
Box 35, FIN-40351 Jyv\"askyl\"a, Finland}
\email{andrei.rybin@phys.jyu.fi}
\author{ I.P. Vadeiko}
\affiliation{School of Physics and Astronomy, University of St
Andrews, North Haugh, St Andrews, KY16 9SS, Scotland}
\email{iv3@st-andrews.ac.uk}

\date{ July 10, 2004}
\begin{abstract}

We propose a   method to control the optical transparency of a
Bose-Einstein condensate with working energy levels of the
$\Lambda$-type. The reported effects are essentially nonlinear and
are considered in the framework of an exactly solvable model
describing the interaction of light with a $\Lambda$-type medium. We
show how the complicated nonlinear interplay between fast and slow
solitons in the $\Lambda$-type medium points to a possibility to
create optical gates as well as to a possibility to store optical
information.
\end{abstract}
\maketitle

\section{\label{sec1:level1}Introduction.}
Recent progress in experimental techniques for the coherent control
of light-matter interaction opens many opportunities for interesting
practical applications. The experiments are carried out on various
types of materials such as cold sodium atoms \cite{Hau:1999,
Liu:2001}, rubidium atom vapors \cite{Phillips:2001, Bajcsy:2003,
Braje:2003, Mikhailov:2004}, solids \cite{Turukhin:2002,
Bigelow:2003}, photonic crystals \cite{Soljacic:2004}. These
experiments are based on the control over the absorption properties
of the medium and study slow-light and superluminal light effects.
The control can be realized in the regime of electromagnetically
induced transparency (EIT), by the coherent population oscillations
or other induced transparency techniques. The use  of each different
materials brings in specific advantages important for the practical
realization of the effects. For instance, the cold atoms have
negligible doppler broadening and small collision rates, which
increases ground-state coherence time. The experiments on rubidium
vapors are carried at room temperatures and this does not require
application of complicated cooling methods. The solids are obviously
one of the strongest candidates for realization of long-living
optical memory. Photonic crystals provide a broad range of ways to
guide and manipulate the slow light.

\begin{figure}
\includegraphics[width=60mm]{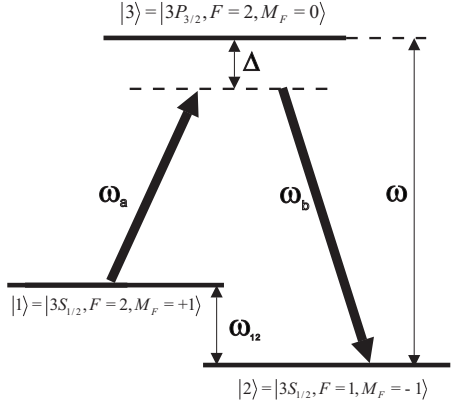}
\caption{\label{fig:spec1} The  $\Lambda$-scheme for working
energy levels of sodium atoms. { The parameters of the scheme are
the following: $\omega_{12}/(2\pi)=1772 \mathrm{MHz}$,
$\omega/(2\pi)=5.1\cdot 10^{14} \mathrm{Hz}$ ($\lambda=589
\mathrm{nm}$), and $\Delta$ is the variable detuning from the
resonance.}}
\end{figure}

In this paper we study the interaction of light with a gaseous
active medium whose working energy levels are well approximated by
the $\Lambda$-scheme. Our theoretical model is a very close
prototype for a gas of sodium atoms, whose interaction with the
light is approximated by the structure of levels of the
$\Lambda$-type. The structure of levels is given in
Fig.~\ref{fig:spec1}, where two hyperfine sub-levels of sodium state
$3^2S_{1/2}$ with $F=1, F=2$ are associated with $|2\rangle$ and
$|1\rangle$ states, { correspondingly~\cite{Hau:1999}}. The excited
state $|3\rangle$ corresponds to the hyperfine sub-level of the term
$3^2P_{3/2}$ with $F=2$. We consider the case when the atoms are
cooled down to microkelvin temperature in order to suppress the
Doppler shift and increase the coherence life-time for the ground
levels. The atomic coherence life-time in sodium atoms at
temperature of $0.9 {\mu}$K is of the order of 0.9 ms
\cite{Liu:2001}. Typically, in the experiments the pulses have
length of microseconds, which is much shorter than the coherence
life-time and longer than the optical relaxation time of $16.3 ns$.
The gas cell is illuminated by two circularly polarized optical
beams co-propagating in the z-direction. One beam, denoted as
channel $a$, is a $\sigma^-$-polarized field, and the other, denoted
as $b$, is a $\sigma^+$-polarized field. The corresponding fields
are presented within the slow varying amplitude and phase
approximation (SVEPA) as
\begin{equation}\label{fields}
     \vec{E}=\vec{e}_a\, \mathcal{E}_a e^{i(k_a z-\omega_a t)} +
     \vec{e}_b\, \mathcal{E}_b e^{i(k_b z-\omega_b t)} +c.c.
\end{equation}
Here, $k_{a,b}$ are the wave numbers, while the vectors
$\vec{e}_a,\vec{e}_b$ describe polarizations of the fields. It is
convenient to introduce two corresponding Rabi frequencies:
\begin{equation}\label{Rabi_f}
    \Omega_a=\frac{2\mu_{a}\mathcal{E}_a}{\hbar},
    \Omega_b=\frac{2\mu_{b}\mathcal{E}_b}{\hbar}.
\end{equation}
Here $\mu_{a,b}$ are dipole moments of quantum transitions in the
channels $a$ and $b$.

Within the SVEPA the Hamiltonian describing the interaction of a
three-level atom with the fields assumes the form
\begin{eqnarray}
\label{H_lam} &H_\Lambda=H_0+H_I,\;H_0=\omega_{12}|1\rangle\langle1|
+\omega|3\rangle\langle3|,\\
&H_I=-\frac12 \left({\Omega_a e^{i(k_a z-\omega_a t)}
|3\rangle\langle1| +\Omega_b e^{i(k_b z-\omega_b t)}
|3\rangle\langle2|}\right) +h.c.\nonumber
\end{eqnarray}
Here $\hbar=1$. The description can be further simplified in the
interaction representation  for the density matrix $\rho$, namely,
$$
\bar\rho=e^{i H_E (t-\frac zc)} \rho e^{-i H_E (t-\frac zc)},$$
$$
H_E=(\frac\Delta 2+\omega_{12})|1\rangle\langle1|+\frac \Delta
2|2\rangle\langle2|+(\omega-\frac \Delta 2)|3\rangle\langle3|.$$
Then the Liouville equation for the transformed operator $\bar\rho$
assumes the form
$$
    i\partial_t \bar \rho=\left[{(H_0-H_E+\bar
    H_I),\bar\rho}\right],
$$
where
\begin{eqnarray}
\label{HI_bar}
   &\bar H_I=-\frac12  \left({\Omega_a |3\rangle\langle1|
  +\Omega_b |3\rangle\langle2|
  }\right)+h.c. =\nonumber\\
  &-\frac12\left(%
\begin{array}{ccc}
  0 & 0 & \Omega_a^* \\
  0 & 0 & \Omega_b^* \\
  \Omega_a & \Omega_b & 0 \\
\end{array}%
\right).
\end{eqnarray}
The dynamics of the fields is described by the Maxwell equations
\begin{eqnarray}\label{Maxwell}
    &(\partial_t^2-c^2 \partial_z^2) \Omega_a e^{i(k_a z-\omega_a
    t)}= -\frac {2 \nu_a}{\omega_a} \partial_t^2 \left({\bar\rho_{31}
    e^{i(k_a z-\omega_a t)}}\right),\nonumber\\
    &(\partial_t^2-c^2 \partial_z^2) \Omega_b e^{i(k_b z-\omega_b
    t)}= -\frac {2 \nu_b}{\omega_b} \partial_t^2 \left({\bar\rho_{32}
    e^{i(k_b z-\omega_b t)}}\right),\nonumber
\end{eqnarray}
where $\nu_a=(n_A |\mu_{a}|^2 \omega_a)/\epsilon_0$, $\nu_b=(n_A
|\mu_{b}|^2 \omega_b)/\epsilon_0$, $n_A$ is the density of atoms,
and $\epsilon_0$ is the vacuum susceptibility. For many experimental
situations it is typical that the coupling constants $\nu_{a,b}$ are
almost the same. Therefore we assume that $\nu_a=\nu_b=\nu_0$.
Within the SVEPA the wave equations are reduced to the first order
PDEs:
\begin{eqnarray}\label{Maxwell_r}
    \partial_\zeta \Omega_a = i \nu_0 \bar\rho_{31}, \;
    \partial_\zeta \Omega_b = i \nu_0 \bar\rho_{32}.\quad
\end{eqnarray}
Here $\zeta=z/c, \tau=t-z/c$.

Equations Eqs.(\ref{Maxwell_r}) can be rewritten in a matrix form as
\begin{equation}\label{Maxwell_m}
    \partial_\zeta \bar H_I=i\frac {\nu_0}4 \left[{D,\bar\rho}\right],
\end{equation}
where
$$
D=\left(%
\begin{array}{ccc}
  1 & 0 & 0 \\
  0 & 1 & 0 \\
  0 & 0 & -1 \\
\end{array}%
\right).
$$
In the new variables the Liouville equation takes the form
\begin{equation}\label{Liouv}
    \partial_\tau \bar \rho=i\left[{\frac\Delta2 D-\bar
    H_I,\bar\rho}\right].
\end{equation}

As was indicated above, the system of equations
Eqs.(\ref{Maxwell_m}),(\ref{Liouv}) is exactly solvable in the
framework of the inverse scattering (IS) method \cite{fad}.

This means that the system of equations
Eqs.(\ref{Maxwell_m}),(\ref{Liouv}) constitutes a compatibility
condition for a certain linear system, namely
\begin{eqnarray}
\label{lin_system}
 &\partial_\tau \Psi= U(\lambda)\,\Psi=\frac i2\lambda D \,\Psi\,  - i \bar H_I
 \,\Psi,\label{lin_system_1}\\
& \partial_\zeta \Psi=
V(\lambda)\,\Psi=\frac{i}{2}\frac{\nu_0\bar\rho}{\lambda-\Delta}
\,\Psi\,.\label{lin_system_2}
\end{eqnarray}
Here, $\lambda\in {\Bbb C}$ is the spectral parameter.    The
comparison $\Psi_{\tau\zeta}$ against $\Psi_{\zeta\tau}$ leads to
the zero-curvature condition \cite{fad}
$U_\zeta(\lambda)-V_\tau(\lambda)+\left[U(\lambda),V(\lambda)\right]=0$,
which holds identically with respect to the linearly independent
terms in $\lambda$. It is straightforward to check that the
resulting conditions coincide with the nonlinear equations
Eqs.(\ref{Maxwell_m}),(\ref{Liouv}). At this point it is worth
discussing the initial and boundary conditions underlying the
physical problem in question. We consider a semi-infinite
$\zeta\ge0$ active medium with a pulse of light
 incident at the point $\zeta=0$ (initial condition). This means
 that the evolution is considered with respect to the {\it space}
 variable $\zeta$, while the boundary conditions should be specified with respect to the  variable $\tau$.
In our case we use as the asymptotic boundary conditions the
asymptotic values of the density matrix as $|\tau|\to\infty$. To
solve the nonlinear dynamics as described by equations
Eqs.(\ref{Maxwell_m}),(\ref{Liouv}), the IS method considers the
scattering problem for the linear system Eq.(\ref{lin_system_1}),
while the auxiliary linear system Eq.(\ref{lin_system_2}) describes
the  evolution of the scattering data. The purpose of this work is,
in particular, to study an essentially nonlinear interplay of the
{fields in the both channels}.  This goal leads to consideration for
equation Eq.(\ref{lin_system_1}) the scattering problem of finite
density type (cf. \cite{fad} and references therein), i.e.
$\Omega_{a,b}\to\Omega_{a,b}^{(0)}$ as $|\tau|\to\infty$.

The scattering  problem for equation Eq.(\ref{lin_system_1}),
reformulated as a matrix Riemann-Hilbert problem, is then posed on
the two-sheet Riemann surface  for the spectral parameter $\lambda$.
The purpose of this work is to discuss certain experimentally
relevant solutions of the $\Lambda$-system, while the technicalities
of the IS method are very well reflected in the existing literature
anyway. For an account of results for   the $\Lambda$-system
accessible through the IS method see, for example, references
\cite{Hioe:1994, Grobe:1994, Park:1998,gab}. We only wish to
indicate that in our analysis we use elements of the approach based
on the Riemann-Hilbert matrix problem, along with algebraic
techniques of the  theory of solitons. We use Darboux-B\"acklund
transformations, in the spirit of \cite{ryb1,ryb2, ryb3, salle}, up
to certain modifications, however (cf. Appendix). The paper is
organized as follows. In the next section we introduce the notions
of slow and fast solitons and describe a nonlinear mechanism of
formation of the transparency gate for the slow soliton. Section III
describes the transparency gate for the fast soliton. The section VI
is devoted to conclusions. The details of the Darboux-B\"acklund
transformation for the $\Lambda$-system are given in the Appendix.

\section{\label{sec2:level1} The transparency gate  for the slow soliton}
In this section we introduce a concept of slow and fast solitons
in the $\Lambda$-medium and explain how the nonlinear interplay
between the solitons leads to  a possibility to control
transparency of the medium. We discuss first the mechanism of
 transparency control  for the slow soliton. We explain how the
fast soliton  propagating in the {\it a} channel hops to the {\it
b} channel where the slow soliton is propagating. The fast soliton
then "knocks down" the slow soliton, thus stopping the propagation
of the latter, and then disappears itself due to the strong
relaxation in the system.

As was indicated above, in this work  we consider exact solutions of
the Maxwell-Bloch system Eqs.(\ref{Maxwell_m}),(\ref{Liouv})
existing on a non-vanishing finite background.  In our
considerations the background field plays the same role as the
controlling field in the conventional linear theory of EIT. The
background field enters the exact solutions as a parameter in a
substantially nonlinear fashion. Because of reasons of experimental
relevance we specify the background field as
\begin{equation}\label{init_fields0}
    \Omega^{(0)}_a=\cos(\eta)\Omega,\; \Omega^{(0)}_b=\sin(\eta)\Omega,\;
\Omega=\Omega_0 e^{i k \zeta}.
\end{equation}
Therefore $\Omega_0^2$ specifies the total intensity of background
fields existing in the both channels, $\eta$ defines the relative
intensities, while $k\ll k_{a,b}$ is introduced in order to take
into account small spatial variations of the phase. The intensity
of the background field  $\Omega_0$ is an experimentally
adjustable parameter.  We show in the paper that this parameter
provides the control over the transparency of the optical gates
and determines the speed of the slow soliton in the system.

For simplicity, we first assume $k=0$. The atoms   are initially
unexcited and are prepared in the so-called dark-state
\begin{equation}\label{dark}
    |\psi_{d}\rangle=\cos(\eta) |2\rangle-\sin(\eta) |1\rangle
\end{equation}
Hereafter we assume $\eta=0$, and therefore the dark state is simply
$|2\rangle$. Taking this state of the atomic subsystem  as the
initial state we find the following solutions describing the
formation of the transparency gate (cf appendix for details):
\begin{widetext}
\begin{eqnarray}\label{Soliton1_1}
    &\tilde\Omega_a=\Omega_0-\frac{2(a_3^2\Omega_0e^{\tau
    \sqrt{\varepsilon_0^2-\Omega_0^2}}+a_2^2\Omega_0e^{-\tau
    \sqrt{\varepsilon_0^2-\Omega_0^2}}+2a_2a_3\varepsilon_0)} {a_3^2e^{\tau
    \sqrt{\varepsilon_0^2-\Omega_0^2}}+a_2^2 e^{-\tau
    \sqrt{\varepsilon_0^2-\Omega_0^2}}+2a_2a_3\frac{\Omega_0}{\varepsilon_0}+
    a_1^2e^{-\tau\varepsilon_0+\zeta\frac{\nu_0\varepsilon_0}
    {\Delta^2+\varepsilon_0^2}}}\nonumber\\
    &\tilde\Omega_b=-2i\frac{\sqrt{2\varepsilon_0} a_1
    e^{-\frac\tau2\varepsilon_0+\zeta\frac{i\nu_0}
    {2(\Delta-i\varepsilon_0)}}(a_3e^{\frac\tau2
    \sqrt{\varepsilon_0^2-\Omega_0^2}} \sqrt{\varepsilon_0+
    \sqrt{\varepsilon_0^2-\Omega_0^2}}+ a_2 e^{-\frac\tau2
    \sqrt{\varepsilon_0^2-\Omega_0^2}}\sqrt{\varepsilon_0-
    \sqrt{\varepsilon_0^2-\Omega_0^2}})} {a_3^2e^{\tau
    \sqrt{\varepsilon_0^2-\Omega_0^2}}+a_2^2 e^{-\tau
    \sqrt{\varepsilon_0^2-\Omega_0^2}}+2a_2a_3\frac{\Omega_0}{\varepsilon_0}+
    a_1^2e^{-\tau\varepsilon_0+\zeta\frac{\nu_0\varepsilon_0}
    {\Delta^2+\varepsilon_0^2}}}.
\end{eqnarray}
\end{widetext}
\noindent The solution Eq.(\ref{Soliton1_1}) describes  the
nonlinear interaction of the slow and fast solitons. As is explained
in the appendix, the solution is parameterized by the constants
defining the position and phase of the solitons. Without loss of
generality we assume $a_1$ and $a_3$ to be real constants, while
$a_2=1$. We show below that $a_1$ determines the position of the
slow soliton whereas $a_3$ determines the position of the fast
signal. In practice these constants are defined by the initial
condition, which specifies  the actual pulse of light entering the
medium at the point $\zeta=0$. To understand the structure of the
slow soliton one can put $a_3=0$. This choice corresponds to taking
the fast soliton to $-\infty$ in the variable $\tau$. Indeed, this
specification removes from the overall solution
Eq.(\ref{Soliton1_1}) the fast pulse component and thus singles out
the slow soliton part. The slow soliton solution assumes then the
following form:
\begin{eqnarray}
    &&\tilde\Omega_a=\Omega_0\tanh(\phi_s)\\
    &&\label{ss_field}\tilde\Omega_b=
    \frac{-ie^{i\frac{\zeta \nu_0 \Delta}{2(\Delta^2+\varepsilon_0^2)}}
    \sqrt{2\varepsilon_0(\varepsilon_0-
    \sqrt{\varepsilon_0^2-\Omega_0^2})}}{\cosh(\phi_s)},
\end{eqnarray}
where
\begin{equation}\label{ss_phase}
\phi_s=\zeta\frac{\varepsilon_0\nu_0}{2(\Delta^2+\varepsilon_0^2)}
-\frac\tau2\left({\varepsilon_0-\sqrt{\varepsilon_0^2-\Omega_0^2}}\right)+\ln|a_1|,
\end{equation}
is the phase of the slow soliton.  From the expression above and in
the simplifying approximation
$\frac{\varepsilon_0^2}{\Omega_0^2}>>1$   the group velocity of the
slow soliton can be easily derived:
$$v_g\approx
c\frac{\Omega_0^2(\Delta^2+\varepsilon_0^2)}{2\varepsilon_0^2\nu_0}.$$

The pure state of the atomic subsystem  corresponding to  the slow
soliton solution Eq.(\ref{ss_field}) reads
\begin{eqnarray}\label{ss_wavefun}
&|\psi\rangle=\left({\frac{\Delta}{\sqrt{\Delta^2+\varepsilon_0^2}}+
i\frac{\varepsilon_0}{\Omega_0}\tilde\Omega_a}\right)|2\rangle\nonumber\\
&+\tilde\Omega_b\frac{i\sqrt{\varepsilon_0+
\sqrt{\varepsilon_0^2-\Omega_0^2}}\;|1\rangle+\sqrt{\varepsilon_0-
    \sqrt{\varepsilon_0^2-\Omega_0^2}}\;|3\rangle}{2\sqrt{\Delta^2+\varepsilon_0^2}\sqrt{\varepsilon_0-
    \sqrt{\varepsilon_0^2-\Omega_0^2}}}
.
\end{eqnarray}
Notice that the population of the upper level $|3\rangle$ is
proportional to the intensity of the background field. The speed of
the slow soliton is also proportional to $\Omega_0^2$. This means
that the slower the soliton, the smaller the population of the level
$|3\rangle$  and, therefore, the dynamics of the  nonlinear system
as a whole is less affected by the relaxation process.

To understand the structure of the fast soliton one can choose
$a_1=0$.  We then arrive at an expression describing a signal moving
on   the constant background  with the speed of light (fast
soliton):
\begin{equation}\label{fs_field}
    \tilde\Omega_a=\Omega_0\left({1-2\frac{\cosh(\phi_f)+\frac{\varepsilon_0}{\Omega_0}}
     {\cosh(\phi_f)+\frac{\Omega_0}{\varepsilon_0}}}\right),\;
    \tilde\Omega_b=0,
\end{equation}
where  the phase of the fast soliton is
$$
\phi_f=\tau\sqrt{\varepsilon_0^2-\Omega_0^2}+\ln|a_3|.
$$
We wish to emphasize once again that the atomic subsystem is
prepared in the dark state $|2\rangle$, which is not affected by a
signal propagating in the channel {\it a}. This explains why the
fast soliton propagates with the  velocity of light.  The fast
signal propagating in channel {\it a} can be observed as a localized
peak in the field intensity $I_a$. Below we analyze how the fast
soliton collides with the slow soliton, hops into channel {\it b},
and then slightly slows down.

\begin{figure}
\includegraphics[width=90mm]{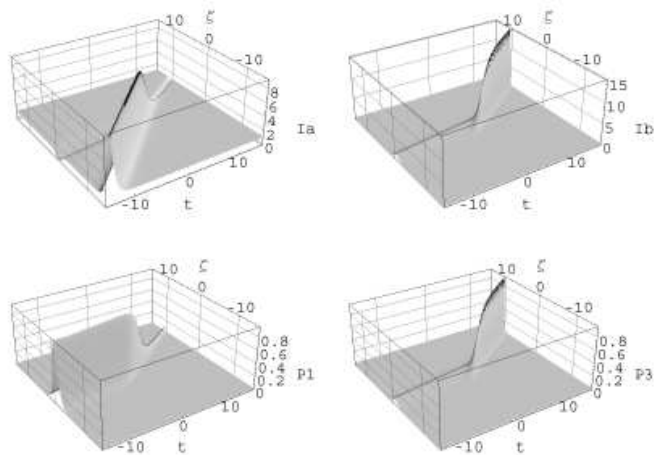}
\caption{\label{fig:spec2} Knocking down the slow soliton. The two
upper plots correspond to the dynamics of the intensities of the
fields $\Omega_a$ and $\Omega_b$. The two lower  plots show the
populations of the levels $|1\rangle$ and $|3\rangle$. The
background field $\Omega_0=1$. Other parameters of plotted solutions
are  always chosen to be the same through the paper, namely
$c_1=c_2=c_3=1, \lambda_0=2i, \nu_0=3, \Delta=0$.}
\end{figure}

Figure~\ref{fig:spec2} illustrates the propagation and collision of
the slow and fast solitons according to equation
Eq.(\ref{Soliton1_1}). The figure for $I_a$ shows the intensities of
the signals in channel $a$. The slow soliton corresponds to a groove
in the background field $\Omega_0$. It is clearly seen that after
the act of collision the slow soliton ceases propagating in channel
$a$, while some trace of the fast soliton still can be noticed in
that channel. The figure for $I_a$ is complemented by the figure for
the intensity $I_b$ of the field in channel $b$. We see that before
the collision only the slow soliton exists in channel $b$, while
after the collision the slow soliton disappears and a  fast
intensive signal appears, whose velocity is slightly below the speed
of light.   The process described above can be summarized as if the
fast soliton knocks down the slow soliton. The notion of a
transparency gate requires the existence of two distinctly different
regimes, which are transparent (open gate), and opaque (closed
gate). In the absence of the fast soliton the gate is open for the
slow soliton. When the fast soliton is present the slow soliton is
destroyed, while the fast intensive signal created after the
collision in channel $b$ is attenuated due to strong relaxation in
the atomic subsystem. The gate thus closes up in the course of the
dynamics due to the relaxation process. To further explain this
process we provide the Fig.~\ref{fig:spec2} plots for populations of
the levels $|1\rangle$ and $|3\rangle$.

Notice that before the collision the population of the upper atom
level $|3\rangle$  is negligible and is approximately given by the
formula for the slow soliton solution Eq.(\ref{ss_wavefun}) (see the
lower right plot of P3). The populations of the lower levels
$|1,2\rangle$ are determined by the slow soliton (see the lower left
plot of P1). Indeed, the fast signal existing in channel $a$ does
not interact with the atoms   because at the onset of the dynamics
their state coincides with the dark state $|2\rangle$.
Figure~\ref{fig:spec2} shows that after the collision the atoms of
the active medium are highly excited and therefore the level
$|3\rangle$ is strongly  populated.  This leads to the fast
attenuation of the speedy intensive signal in channel $b$ due to the
relaxation. The optical gate closes up.

To this point we have described a mechanism of controlling  the
transparency of the medium for a particular type of slowly moving
signals. We now  discuss a possibility to read information stored in
the atomic subsystem.   Let us  assume that the background field
vanishes, i.e. $\Omega_0=0$. As was explained above, the speed of
the slow soliton then vanishes as well. However, the  information
about polarization of the slow signal is stored  in the atomic
subsystem. This effect can be interpreted in terms of the concept of
a polariton, which is a collective excitation  of the overall
atom-field system. The notion of a polariton for the
$\Lambda$-system has been used before. In the linear case the
dark-state polariton was discussed in \cite{Fleischhauer:2000}. In
the strongly nonlinear regime, which is the case for the present
work,  a similar interpretation is possible. Indeed, the field
component of a slow soliton solution can be interpreted as the light
contribution into the slow polariton. When the controlling field
$\Omega_0$ vanishes this contribution vanishes as well, along with
the speed of the polariton. The latter then contains only
excitations in the atomic subsystem. The general solution
Eq.(\ref{Soliton1_1}) is then reduced to the form:

\begin{eqnarray}\label{zfields_field}
 & \tilde\Omega_a=\frac{-4i c_1 c_3 \varepsilon_0
 \exp\left[{\varepsilon_0\tau-\frac{\zeta\varepsilon_0\nu_0}
     {2(\Delta^2+ \varepsilon_0^2)}}\right]}
     {2c_2 c_3\cosh(\phi_{s0}) + c_1^2
     \exp\left[{2\varepsilon_0\tau-\frac{\zeta\varepsilon_0\nu_0}
     {2(\Delta^2+ \varepsilon_0^2)}}\right]},\nonumber\\
  &  \tilde\Omega_b=\frac{c_2}{c_3}
  e^{\frac{i\zeta\nu_0}
     {2(\Delta+i \varepsilon_0)}}
     \tilde\Omega_a,
\end{eqnarray}
where
$\phi_{s0}=\zeta\frac{\varepsilon_0\nu_0}{2(\Delta^2+\varepsilon_0^2)}+
\ln(c_2/c_3)$ is the phase of the slow soliton for the vanishing
background $\Omega_0$. The form of the fields resembles a
superposition of fast and slow solitons in Eq.(\ref{Soliton1_1}),
with  the vanishing velocity of the slow soliton. It is impossible
to single out the contribution of the slow pulse by choosing
$c_3=0$, because the amplitude of the slow signal approaches zero as
the background field vanishes(see Eq.(\ref{ss_field})). The atomic
state describing the stored information reads
\begin{eqnarray}\label{ss0_wavefun}
  & |\psi\rangle=\sqrt{\frac{\Delta+i\varepsilon_0}{\Delta-i
   \varepsilon_0}}\;|2\rangle  \\
&+\frac{i\tilde\Omega_b}{2\sqrt{\Delta^2+
\varepsilon_0^2}}\left({\frac{c_3}{c_1}e^{-\varepsilon_0
\tau}\;|1\rangle+\frac{c_2}{c_1}e^{-\varepsilon_0
\tau-\frac{i\zeta\nu_0}{2(\Delta-i\varepsilon_0)}}\;|2\rangle
+|3\rangle}\right) .\nonumber
\end{eqnarray}
For $c_1=0$ the fields vanish, while the atomic state reduces to a
form corresponding to a stopped polariton described by
Eq.(\ref{ss_wavefun}) with $\Omega_0=0$. In other words, when the
slow soliton is completely stopped its information is stored in the
spin polarization of the atoms. As long as the upper state
$|3\rangle$ is not populated, the state of the atomic subsystem is
not sensitive to the destructive influence of the optical relaxation
processes.

\begin{figure}
\includegraphics[width=90mm]{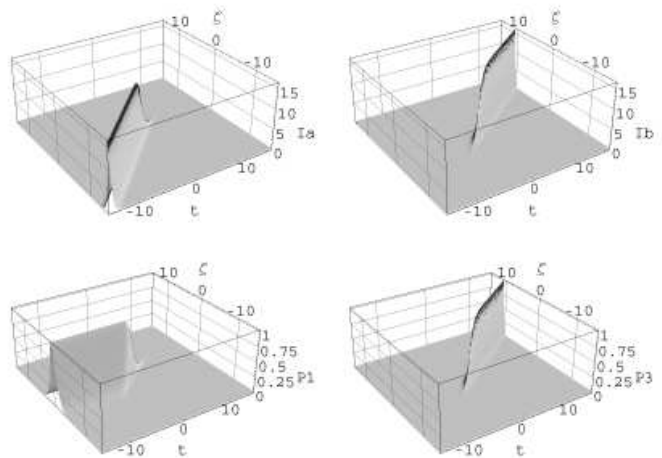}
\caption{\label{fig:spec3} Reading  the optical information by the
fast soliton. The two upper plots illustrate the dynamics  of the
fields $\Omega_a$ and $\Omega_b$. The  two lower  plots show the
populations of the levels $|1\rangle$ and $|3\rangle$. The
standing peak on the plot for $P_1$ corresponds to the stored
information in the form of the  localized polarization. The
rapidly moving localized excitation of the atoms given on the plot
for $P_3$ represents the act of reading. The background field
$\Omega_0=0$.}
\end{figure}

The  conventional way \cite{Liu:2001} to read the information stored
in the atoms is to  increase the intensity of the background field.
Our method of reading the information is different. We propose to
send the fast soliton into the space domain in the active medium,
where the information is stored.  The polarization in the domain is
then flipped by the fast signal. This is how the reading of
information is realized.  This way of reading optical information is
advantageous  because  it involves fast easily detectable processes.
Figure~\ref{fig:spec3} illustrates  the mechanism of the reading.
Notice that the  act of reading, based on the polarization flipping
induced by the fast signal, can be realized in a very short time
scale compared to typical relaxation times.

\section{\label{sec3:level1} The transparency gate for the fast
signal}

\begin{figure}
\includegraphics[width=60mm]{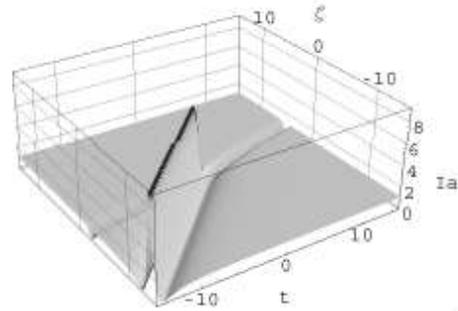}
\caption{\label{fig:spec4} Knocking down the fast signal. The small
localized polarization flip is the slow soliton part. The intensive
peak is the fast signal. The background field $\Omega_0=1,
\lambda_0=i$.}
\end{figure}

In this section we explain how to create the transparency gate for
the fast signal rather than for the slow soliton as in the preceding
section. Instead of decreasing the background intensity we now
propose to increase it to reach the  point where
$\Omega_0=\varepsilon_0$. As we show below (cf Fig.~\ref{fig:spec4}
as well), the fast signal can be stopped by the slow soliton. In
other words, the control of the transparency for the fast signal is
realized by tuning the background intensity to the parameter
$\varepsilon_0$ entering the slow soliton solution. This effect is
related to the appearance of the so-called exulton type solutions
reported before \cite{ryb1}, \cite{ryb5} for the SIT and NLS models.

For the constant background field with $k=0$ and for
$\Omega_0=\varepsilon_0$ the solution has a simple and transparent
form, namely
\begin{eqnarray}\label{exulton_field}
 & \tilde\Omega_a=\Omega_0\frac{c_1^2 e^{\phi_{s1}}-2e^{-\phi_{s1}}
 \left({(c_2+c_3\tau\Omega_0)^2-3c_3^2}\right)}
     {c_1^2 e^{\phi_{s1}}+2e^{-\phi_{s1}}
 \left({(c_2+c_3\tau\Omega_0)^2+c_3^2}\right)},\nonumber\\
  &  \tilde\Omega_b=-\Omega_0\frac{4ic_1 e^{\frac{i\zeta\nu_0\Delta}
  {2(\Omega_0^2+\Delta^2)}}
  \left({c_2+c_3(1+\tau\Omega_0)}\right)}
     {c_1^2 e^{\phi_{s1}}+2e^{-\phi_{s1}}
 \left({(c_2+c_3\tau\Omega_0)^2+c_3^2}\right)},
\end{eqnarray}
and
\begin{eqnarray}\label{exulton_wavefun}
  & |\psi\rangle=\sqrt{\frac{\Delta+i\Omega_0}{\Delta-i
   \Omega_0}}\;|2\rangle  +\frac{\tilde\Omega_b}{2\sqrt{\Delta^2+
\Omega_0^2}}\\
&\left({i\frac{c_2-c_3+c_3 \tau \Omega_0} {c_2+c_3+c_3 \tau
\Omega_0} \;|1\rangle+\frac{c_1 e^{-\Omega_0
\tau-\frac{i\zeta\nu_0}{2(\Delta-i\Omega_0)}}}{c_2+c_3+c_3 \tau
\Omega_0}\;|2\rangle +|3\rangle}\right) .\nonumber
\end{eqnarray}
Here, the phase $\phi_{s1}$ corresponds to $\phi_{s}$ with
$\varepsilon_0=\Omega_0$.

To explain the mechanism underlying the functioning of an optical
gate for the fast signal we analyze the solution
Eq.(\ref{exulton_field}) in detail. In the case $c_3=0$ we obtain
the conventional slow soliton solution. If $c_1=0$   we arrive at a
rational algebraic expression, which is a solution of the exulton
type.  In the case $k=0$ and $c_1=0$ the  pulse, whose algebraic
form is a rational expression (exulton), moves with the speed of
light without decay.

It can be readily seen  that if $k$ is not zero the exulton part of
the solution vanishes in the course of the dynamics. For simplicity,
and to demonstrate the typical behavior of this kind, we choose
$c_1=0, c_2=0$. Then,
\begin{widetext}
\begin{eqnarray}\label{exulton0_field}
 \tilde\Omega_a=\Omega_0\left({1-2e^{ik\zeta}
 \frac{(i\Delta(1+\tau\Omega_0)+\Omega_0(1-ik\zeta+\tau\Omega_0))
 (i\Delta(1-\tau\Omega_0)+\Omega_0(-1+ik\zeta+\tau\Omega_0))}
     {\Delta^2+(1+(k\zeta-\Delta\tau)^2)\Omega_0^2+\tau^2\Omega_0^4}}\right),
     \;\tilde\Omega_b=0.
\end{eqnarray}
\end{widetext}

The collision of the fast signal (exulton) with the slow soliton
results in the disappearance of the former in channel $a$. This
means that the slow soliton controls the transparency of the medium
for the signal moving with the speed of light. The reported
mechanism is a realization of an optical gate for the fast signal.

\section{\label{sec4:level1} Conclusions and discussion }
In this paper we discussed a concept of the transparency gate for
the fast and slow solitons in a $\Lambda$-type medium. We explained
how the fast soliton can knock
 down the slow soliton and close the gate for the latter. The
 opposite process of closing the gate for the fast signal, given
 by rational algebraic expression (exulton), is also described. We
 also described the process of reading optical information,
written into the active medium by the slow soliton. It is worth
discussing here a possibility to actually create in the
$\Lambda$-type atomic medium the signals described above. The
general physical feature underlying  the mathematical property of
complete integrability is a  delicate balance between the
 workings of dispersion and nonlinearity  inherent in the medium.
 Provided that this balance is observed and the system is
 completely integrable it is a general fact that virtually any
 sufficiently intensive localized initial condition creates solitons. The
 overall picture of nonlinear dynamics can be roughly described as
 follows. The evolving signal created by the incident pulse in the
 course of the dynamics breaks down to a number of solitons and a
 decaying tale. The latter vanishes in due course. The soliton-like signals survive
(ideally, i.e. in the absence of dissipation) for infinitely long
time.   When the physical conditions underlying the complete
integrability of the optical  system are met, the general picture of
the nonlinear dynamics is  similar to the described above. Namely, a
fairly arbitrary localized  and intensive initial signal creates in
the  course of nonlinear dynamics a number of solitons.  The number
of  solitons can be  derived from the analysis of the corresponding
zero-curvature  representation. The signal Eq.(\ref{Soliton1_1}) is
a  rather generic soliton-like solution of the nonlinear system
 Eqs.(\ref{Maxwell_m}), (\ref{Liouv}) and therefore it is very
 plausible that such signal can be created. Of course an
 experiment accompanied by a mathematical scrutiny would be
 necessary to further support this claim. The purpose of
this paper is merely limited to pointing to a possibility of
interesting  physical applications. We want to emphasize that in our
considerations the distinguished role  is assigned to the background
field $\Omega_0$ that turns out to be a nonlinear analog of the
conventional controlling field appearing in the linear EIT
formulation. The difference between the linear and nonlinear cases
lies in the fact that in the nonlinear case the control field and
the soliton solution are present in the same channel in inseparable
fashion of nonlinear superposition. In our future work (ArXiv:
quant-ph/0411148; quant-ph/0411149) we plan to further support our
results on memory reading and transparency control by an
investigation of transient regimes, when the background field
$\Omega_0$ adiabatically vanishes or adiabatically approaches the
point $\epsilon_0$. The treatment of relaxation processes in the
system will  be also included. This forthcoming investigation will
use the method of collective variable \cite{bish} along with other
perturbation techniques developed for the systems integrable by the
inverse scattering method.

\section*{Acknowledgements}
AR wishes to thank for kind hospitality Professor Ulf Leonhardt and
the University of St. Andrews, Scotland, where a part of this work
was carried out. IPV acknowledges the support of the Engineering and
Physical Sciences Research Council.

\bibliography{paper_actual}

\section{\label{sec5:level1}Appendix. Darboux-B\"acklund transformation for
the $\Lambda$-system}

In this appendix we describe the Darboux-B\"acklund (DB)
transformation for the $\Lambda$-system. It is plain to see that the
linear system Eqs.(\ref{lin_system_1}),(\ref{lin_system_2}) is
covariant with respect to the following DB dressing transformation
\begin{eqnarray}\label{Darboux_1}
&\tilde H_I=\bar H_I -\frac12 \left[{D,\sigma_1(0)}\right],\; \tilde
\rho=\sigma_1(\Delta)\, \bar\rho\, \sigma_1^{-1}(\Delta) \\
& \tilde\Psi=\Psi{\cal L}-\sigma_1(0)\Psi
,\;\sigma_1(\Delta)=\Psi_1\left({{\cal L}_1-\Delta}\right)
\Psi_1^{-1}\label{Darboux_2}.
\end{eqnarray}
Here $\Psi$ is a matrix consisting of three linearly independent
solutions of the linear system
Eqs.(\ref{lin_system_1}),(\ref{lin_system_2}) corresponding to three
(not necessarily different) values of the spectral parameter
$\lambda$. The matrix spectral parameter ${\cal L}$ in our case is
defined as
$$
{\cal L}=\left(%
\begin{array}{ccc}
  \lambda' & 0 & 0 \\
  0 & \lambda''& 0 \\
  0 & 0 & \lambda''' \\
\end{array}%
\right)_,
$$
where $\lambda', \lambda'', \lambda'''$ are certain values of the
spectral parameter $\lambda$. The matrix $\Psi_1$ is a specification
of $\Psi$ corresponding to the following particular value of the
matrix spectral parameter:

$$
{\cal L}_1=\left(%
\begin{array}{ccc}
  \lambda_0^* & 0 & 0 \\
  0 & \lambda_0^*& 0 \\
  0 & 0 & \lambda_0 \\
\end{array}%
\right).
$$
We denote the fundamental matrix of the linear system
Eq.(\ref{lin_system}) for $\lambda=\lambda_0$ as $\Phi_0$.

It can be shown that for the value of the spectral parameter
$\lambda=\lambda_0^*$ the fundamental matrix is
$\bar{\Phi}_0\equiv(\Phi_0^{-1})^\dag$. Since the subspace of
solutions corresponding to $\lambda_0^*$ is two dimensional, the
matrix $\Psi_1=(\vec{\psi}_1,\vec{\psi}_2,\vec{\psi}_3)$ is
constructed as follows. The vector $\vec{\psi}_3=c_1
\Phi_0^{(1)}+c_2 \Phi_0^{(2)}+c_3 \Phi_0^{(3)}$ is a general
solution of the linear problem with $\lambda=\lambda_0$. Here
$\Phi_0^{(i)},\,i=1,2,3$ denotes a column in the matrix $\Phi_0$. To
satisfy the structure of the operator $\sigma_1$
Eq.(\ref{Darboux_2}) we  require that $(\vec{\psi}_3,
\vec{\psi}_{1(2)})=0$ and the vectors $\vec{\psi}_{1(2)}$ correspond
to $\lambda=\lambda_0^*$. Here, $(\cdot ,\cdot )$ denotes a scalar
product of two vectors in 3D complex vector space. It follows from
the definition of the matrix $\bar{\Phi}_0$ that
$(\bar{\Phi}_0^{(i)}, \Phi_0^{(j)})=\delta_{i,j}$. Therefore, we can
easily find two appropriate orthogonal vectors $\vec{\psi}_{1,2}$:
$$\vec{\psi}_1=
(c_2^*+ c_3^*) \bar{\Phi}_0^{(1)}-c_1^* ( \bar{\Phi}_0^{(2)}+
\bar{\Phi}_0^{(3)});\; \vec{\psi}_2=c_3^* \bar{\Phi}_0^{(2)}-c_2^*
\bar{\Phi}_0^{(3)}.$$

The algorithm of finding new solutions of the nonlinear system
Eqs.(\ref{Maxwell_m}),(\ref{Liouv}) can be recapitulated as
follows. Find a solution $\Phi_0$ of the associated linear system
Eqs.(\ref{lin_system_1}),(\ref{lin_system_2}), corresponding to a
certain "seed" solution of the nonlinear system
Eqs.(\ref{Maxwell_m}),(\ref{Liouv}). Build $\Psi_1$, build
$\sigma_1$, use then the dressing transformation
Eq.(\ref{Darboux_1}).

In the case considered in the present work the  matrix $\Phi_0$
reads
\begin{eqnarray}\label{Phi_0}
&\Phi_0=\left(%
\begin{array}{ccc}
  -\textrm{tan}(\eta) & \frac{\Omega_0 \cos(\eta)}
  {-\lambda_0+i\sqrt{-\lambda_0^2-\Omega_0^2}} & -\frac{\Omega_0 \cos(\eta)}
  {\lambda_0+i\sqrt{-\lambda_0^2-\Omega_0^2}} \\
  1 & \frac{\Omega_0 \sin(\eta)}
  {-\lambda_0+i\sqrt{-\lambda_0^2-\Omega_0^2}} & -\frac{\Omega_0 \sin(\eta)}
  {\lambda_0+i\sqrt{-\lambda_0^2-\Omega_0^2}} \\
  0 & 1 & 1 \\
\end{array}%
\right)\nonumber\\
&e^{i\frac{k\Delta\zeta}{2(\lambda_0-\Delta)}}\left(%
\begin{array}{ccc}
  \exp(\mu_1) & 0 & 0 \\
  0 & \exp(-\mu_2) & 0 \\
  0 & 0 & \exp(\mu_2) \\
\end{array}%
\right),
\end{eqnarray}
where
\begin{eqnarray}
&\mu_1=i\frac{\lambda_0}2\left({\tau+\frac{k\zeta}
{\lambda_0-\Delta}}\right)+i\frac{\nu_0-4k\Delta}
{2(\lambda_0-\Delta)} \zeta,\nonumber\\
&\mu_2=\frac{\sqrt{-\lambda_0^2-\Omega_0^2}}2\left({\tau+\frac{k\zeta}
{\lambda_0-\Delta}}\right).\nonumber\end{eqnarray}

The constants $c_{1,2,3}$ above define the spatial position and
phase of the slow and fast solitons at a fixed moment of time. For
simplicity, we assume  these constants to be real.

To access the solution Eq.(\ref{Soliton1_1}) one has to specify
$c_1=a_1 \Omega_0\sqrt{\frac{\varepsilon_0}
{2(\varepsilon_0^2-\Omega_0^2)}}$, $c_2=a_2\sqrt{\varepsilon_0-
    \sqrt{\varepsilon_0^2-\Omega_0^2}}$, and
    $c_3=a_3\sqrt{\varepsilon_0+
    \sqrt{\varepsilon_0^2-\Omega_0^2}}$, while $\lambda=i\varepsilon_0$.
 This concludes the construction of the
Darboux-B\"acklund dressing transformation for the   nonlinear
$\Lambda$-model.

\end{document}